\documentstyle[editedvolume,numreferences]{crckapb} 

\def\spose#1{\hbox to 0pt{#1\hss}}
\def\lta{\mathrel{\spose{\lower 3pt\hbox{$\mathchar"218$}}
     \raise 2.0pt\hbox{$\mathchar"13C$}}}
\def\gta{\mathrel{\spose{\lower 3pt\hbox{$\mathchar"218$}}
    \raise 2.0pt\hbox{$\mathchar"13E$}}}

           \def\mx{m_{\rm x}}        \def\mc{m_{\rm c}}
 \def\mP{m_{_{\rm P}}}         
 \def\Ev{E_{\rm v}}         \def\lv{\ell_{\rm v}}     \def\Rv{R_{\rm v}}
          \def\nv{n_{\rm v}}        
 \def\nua{\nu_\star}        
 \def\rhv{\rho_{\rm v}}     \def\rhb{\rho_{\rm b}}    
 \def\rhc{\rho_{\rm c}}     \def\rhN{\rho_{_{\rm N}}}
 \def\aa{{g^*}}                     

 \def\Tx{\Th_{\rm x}}

 \def\TN{\Th_{_{\rm N}}}

\def\Journal#1#2#3#4{{#1} {\bf #2}, #3 (#4)}
\def\AP{\em Ann. Phys.}

\def\NPB{{\em Nucl. Phys.} B}
\def\PLB{{\em Phys. Lett.}  B}
\def\PRL{\em Phys. Rev. Lett.}
\def\PRD{{\em Phys. Rev.} D}

\def\NCA{\em Nuovo Cimento A}
\def\be{\begin{equation}}
\def\ee{\end{equation}}
\def\bea{\begin{eqnarray}}
\def\eea{\end{eqnarray}}

      \def\Ov{{\mit\Omega}_{\rm v}}

      \def\GeV{\,{\rm GeV}}

           \def\mx{m_{\rm x}}        \def\mc{m_{\rm c}}
   
 \def\Ev{E_{\rm v}}         \def\lv{\ell_{\rm v}}     \def\Rv{R_{\rm v}}
          \def\nv{n_{\rm v}}        
 \def\nua{\nu_\star}        
 \def\rhv{\rho_{\rm v}}     \def\rhb{\rho_{\rm b}}    
 \def\rhc{\rho_{\rm c}}     \def\rhN{\rho_{_{\rm N}}}
 \def\aa{{g^*}}                     
 \def\aas{{g^*_s}}

 \def\Tx{T_{\rm x}}

 \def\TN{T_{_{\rm N}}}

\newcommand{\dGS}{\delta_{_{GS}}}
\newcommand{\vp}{\varphi}

\newcommand{\sR}{s_{\!\scriptscriptstyle R}}

\newcommand{\g}{\gamma}

\newcommand{\bref}[1]{(\ref{#1})}
\newcommand{\Lag}{{\cal L}}

\newcommand{\al}{\alpha}
\newcommand{\ald}{{\dot{\alpha}}}
\newcommand{\sm}{{\sigma^\mu}}

\newcommand{\sgth}{\sigma^\vp}
\newcommand{\sgr}{\sigma^r}
\newcommand{\sgz}{\sigma^z}
\newcommand{\la}{\lambda}
\newcommand{\lab}{\bar{\lambda}}
\newcommand{\psb}{\bar{\psi}}
\newcommand{\phb}{\bar{\phi}}
\newcommand{\xib}{\bar{\xi}}

\newcommand{\dmu}{{\partial_\mu}}
\newcommand{\dth}{{\partial_\vp}}
\newcommand{\dr}{{\partial_r}}
\newcommand{\eth}[1]{e^{{#1}\vp}}
\newcommand{\spinU}{\mbox{\scriptsize 
		$\left(\begin{array}{c} 1 \\ 0 \end{array} \right)$}}

\newcommand{\mat}[4]{\left(\begin{array}{cc} 
			{#1} & {#2} \\ {#3} & {#4} \end{array}\right)}

\begin{opening}
\title{Cosmic Defects and Particle Physics Constraints}

\author{A.C. Davis}
\institute{Department of Applied Mathematics and Theoretical Physics,\\
           CMS, University of Cambridge,\\ 
           Wilberforce Road, Cambridge, CB3 0WA, UK}

\end{opening}

\runningtitle{Cosmic Defects and Particle Physics Constraints}

\begin{document}

\section{Introduction}

Topological defects, and in particular cosmic strings, are natural
consequences of phase transitions in the early universe \cite{VS,HK,Kibb}.
If detected
they could provide windows into physics at very early times
and very high energy, giving important information both for particle
physics and cosmology. For most cosmological
studies the abelian Higgs model is used as a prototypical cosmic string
theory. 
However, in realistic particle physics theories the situation
is more complicated and the resulting cosmic strings can have a rich 
microstructure. Additional features can be acquired at the string core at
each subsequent symmetry breaking. This additional microstructure can, in
some cases, be used to constrain the underlying particle physics theory
to ensure consistency with standard cosmology. For example, if fermions
couple to the Higgs field which gives rise to the string then these
fermions could become zero modes (zero energy solutions of the Dirac
equation) in the string core. 
The existence of fermion zero modes in the string core can have dramatic
consequences for the properties of cosmic strings. For example, 
the zero modes can be excited and will move up or down the string, depending 
on whether 
they are left- or right-movers. This will result in the string carrying 
a current \cite{witten}. An intially weak current on a string loop will be 
amplified as the loop contracts and could become sufficiently strong 
and prevent the string loop from decaying. In this case a stable loop
or vorton \cite{vorton} is produced. 
The density of vortons 
is tightly constrained by cosmology. For example, if vortons 
are sufficiently stable so that they survive until the present time then 
we require that the universe is not vorton dominated. However, if vortons
only survive a few minutes then they can still have cosmological 
implications. We then require that the universe be radiation dominated
at nucleosynthesis. 
These requirements have been used in \cite{rob&brandon,CarterDavis:99} 
to constrain such models. 

Vortons are classically stable 
\cite{vortonstab}. If vortons do decay, then  they probably do so by quantum 
mechanical tunnelling \cite{rick,blanco}. This would result in them being very 
long lived. In the case of fermion zero modes it has been shown that they are 
indeed very stable, particularly in the chiral limit \cite{DDP}.
However, in the case of fermion superconductivity, the existence of fermion 
zero modes at high energy does not guarantee that such modes survive subsequent
phase transitions. It is thus necessary
to trace the microphysics of the cosmic string from formation through all
subsequent phase transitions in the history of the universe. 

For example, many popular particle physics theories above the electroweak scale
are based on supersymmetry. Such theories can also admit cosmic string 
solutions \cite{mark1}. Since supersymmetry is a natural symmetry between
bosons and fermions, the fermion partner of the Higgs field forming the
cosmic string is a zero mode. Thus, the particle content and interactions
dictated by supersymmetry naturally give rise to current-carrying strings.
Gauge symmetry breaking can arise either by introduction of a super-potential
or by means of a Fayet-Iliopoulos term \cite{Fayet II}. In both cases 
fermion zero modes arise.

However, supersymmetry is not observed in nature and must therefore be
broken. General soft supersymmetry breaking terms are introduced and
their effect on the fermion zero modes considered \cite{mark2}.
For most soft breaking terms, the zero modes are destroyed. Hence, any
vortons formed would dissipate. However, in the case of gauge symmetry
breaking via a Fayet-Iliopoulos term \cite{Fayet II}, the zero modes, and 
hence vortons,
survive supersymmetry breaking. Hence, supersymmetric theories which break
a $U(1)$ symmetry in this way would result in cosmologically stable vortons
and could therefore be ruled out. 

In these lectures I review the necessary properties of cosmic strings and 
will introduce current-carrying strings. The formation and properties
of cosmic vortons is discussed and I derive constraints on the underlying
particle physics theories. Cosmic strings in supersymmetric
theories are analysed, and it is shown how they become current-carrying.
The vorton constraints on such theories is considered. I will discuss 
how the observed microwave background radiation can be used to constrain 
particle physics theories which have some form of long lived, but
not absolutely stable defect. Finally I address the problem of
cosmic strings arising in superstring theory. Superstrings introduce
an extra ingredient, the dilaton field. Since dilatons are long-lived
their production result in constraints on the underlying theory. 
These are discussed in an effective field theory.

\section{Cosmic Strings and Current-Carrying Strings}

In this section I am going to summarise briefly the salient features
of cosmic strings and current-carrying strings. The former have been
discussed in previous lectures \cite{tom,mairi} but
I will include a summary for completeness. Further detail can be found
in the above lectures or in the excellent reviews \cite{VS,HK,Kibb}.

Topological defects can be formed when the universe undergoes a phase 
transition. In particular, if the vacuum manifold is topologically
non-trivial then defects result. If the first homotopy group is
not trivial then cosmic strings are formed. The simplest model for
cosmic strings is the abelian Higgs model with lagrangian

\be 
\Lag = D_\mu \phb D^{\mu} \phi
		- \frac{1}{4} F^{\mu\nu}F_{\mu\nu} \ - V(\phi),
\ee
where 
\be
V(\phi) = \lambda(\phb\phi - \eta^2).
\ee
By solving the classical equations of motion one finds the string solution,
\be
\phi = \eta f(r) e^{i\theta},
\ee
where $f(r)$ is zero for small $r$ and tends to $1$ for 
$r\sim{\eta\sqrt{\lambda}}^{-1}$.
This defines the core of the string. Note that in the string core the 
$U(1)$ symmetry is restored, but is broken outside the core. For topological
reasons cosmic strings are either infinite or closed loops.
Immediately
after the phase transition the string network interacts strongly with
the background plasma. This interaction results in the correlation length,
$\xi(t)$, growing until it catches up with the horizon, after which the
correlation length scales with the horizon. Initially the correlation
length is microphysical, being $\xi(t_{\rm x}) \approx \eta^{-1}$, where 
$t_{\rm x}$ is the time of the phase transition producing the cosmic string 
network. By calculating the frictional forces, which
are dominated by Aharonov--Bohm scattering, one finds that $\xi(t)$
grows like $t^{5/4}$ at early times. This is the friction dominated
regime. Eventually the correlation length catches up with the horizon
at time $t_{\star} = (G\mu)^{-1} t_{\rm x}$, where $\mu$ is the string tension
and is approximately $\eta^2$. After $t_{\star}$ the correlation length
scales with the horizon; this is denoted the scaling regime. The scaling
solution is maintained by the interactions of cosmic strings. That is,
when two infinite strings meet they intercommute. Similarly, when an
infinite string self-intersects a closed loop and infinite string result.
Whilst infinite strings are stable, loops oscillate, lose
energy via gravitational radiation and eventually decay. Thus the string
network does not dominate the energy density of the universe. 

However, in realistic models of particle physics the Higgs field also couples
to fermions. Indeed, in the standard model the fermions acquire their
masses by such a Yukawa coupling. This gives rise to a richer
structure of cosmic strings. For example, consider the simple model

\be
\Lag = \psb i\gamma_{\mu}D^{\mu}\psi + \lambda_f\psb_L \phi\psi_R + h.c.,
\ee 
where $\lambda_f$ is the Yukawa coupling. Thus, when $\phi$ goes to zero in the
core of the string, the fermions become massless; they are zero mode
solutions. They can be excited and move up or down the string depending
on their charges and whether they are left- or right-movers. This results in 
the string carrying a current, though not necessarily an electromagnetic 
current. Nevertheless, this is not the most general case. When the
string has bosonic zero modes, new fermion zero modes have been found 
\cite{jon&mike} which acquire mass from Yukawa couplings to scalar fields
that are non-zero in the core.

For current-carrying strings the above evolution of the cosmic string
network can be altered since string loops can be stabilised by the
angular momentum of the current-carriers. This results in stable loops
or vortons and there is the possibility of the string network dominating
the energy density of the universe. In this case constraints can be
put on the underlying particle physics theory. We consider this in the
next section. In the case of electromagnetically charged currents the
frictional effects are also different from that described above. This is
outside the scope or these lectures; further details are in \cite{friction}.

\section{Cosmic Vortons and Particle Physics Constraints}

When a string becomes current-carrying the properties of the network
are modified. In this section we consider the constraints on the
underlying theory when stable loops, or vortons, result.
When the string acquires a current as a 
consequence of fermion zero modes (as in theories when the fermions
become massive from the string-forming Higgs field) the zero modes
will be present in the string core at formation. If we call
the temperature of the phase transition forming the strings $\Tx$, we can
estimate the vorton density. The more general case to consider would be
when the string becomes current-carrying at a subsequent phase transition,
but this is beyond the scope of these lectures; we refer the reader to 
\cite{rob&brandon,CarterDavis:99}. The physics of cosmic vortons has 
also been discussed in \cite{martins,yves}.

The string loop is characterised by two currents, the topologically conserved
phase current and the dynamically conserved particle number current. Thus
the string carries two conserved quantum numbers; $N$ is the topologically
conserved integral of the phase current and $Z$ is the particle number.
A non conducting Kibble type string loop must
ultimately decay by radiative and frictional drag processes until it
disappears completely. However, a conducting
string loop may be saved from
disappearance by reaching a state in which the energy attains a minimum for
given non-zero values of $N$ and $Z$, ie the loop is stabilised by the
angular momentum of the current-carriers. Here we are going to consider
two cases; the chiral and non-chiral case. In the former there are 
either left- or right-movers in the string core, but not both. 

It should be emphasised that the existence of such vorton states does not
require that the carrier field be gauge coupled. If there is indeed a 
non-zero charge coupling then the loop will have a corresponding total
electric charge, $Q$, such that the particle number is $Z=Q/e$. However,
the important point is that, even in the uncoupled case where $Q$ vanishes,
the particle number $Z$ is perfectly well defined. Indeed, in the strictly
chiral case the current is not electromagnetically coupled. 

The physical properties of a vorton state are determined by the quantum 
numbers, $N$ and $Z$. However, these are not arbitrary. For example,
to avoid decaying completely like a non-conducting loop, a conducting loop 
must have a non-zero value for at least one of the numbers $N$ and $Z$. In 
fact, one would expect that both these numbers be reasonably large 
compared with unity. There is a further restriction on the 
values of their ratio $Z/N$ in order to avoid spontaneous 
particle emission as a result of current saturation. In general we would
expect that $\vert Z\vert \approx N $; in the chiral case there is only
one independent quantum number and $\vert Z\vert = N $. The energy of the
vorton is

\be
\Ev\simeq \lv \mx^{\,2}\ \ ,
\ee
where $\mx$ is related to the string tension. 

In order to evaluate this quantity all that remains is to work out $\lv$. 
Assuming that vortons are approximately circular, with radius
given by $\Rv=\lv/2\pi$ and angular momentum
quantum number $J$ given by $J=NZ$ \cite{C ring} one obtains 

\be
\lv\simeq(2\pi)^{1/2}\vert NZ\vert^{1/2}\mx^{-1} \ . 
\ee
Thus we obtain an estimate of the vorton mass energy as
\be
\Ev\simeq(2\pi)^{1/2}\vert NZ\vert^{1/2}\mx\approx N\mx\ ,
\label {energy}
\ee
where we are assuming the classical description of the string dynamics.
This is valid only if the length $\lv$ is large
compared with the relevant quantum wavelengths. This will only be satisfied 
if the product of the quantum numbers $N$ and $Z$ is
sufficiently large. A loop that does not satisfy this requirement will never
stabilise as a vorton. 

We can now calculate the vorton abundance. Assuming that the string becomes
current carrying at a scale $\Tx$ by fermion zero modes then
one expects that thermal fluctuations will give rise to a non-zero value for
the current. Hence, a random walk process
will result in a spectrum of finite values for the corresponding
string loop quantum numbers $N$ and $Z$. Therefore, loops for which these
numbers satisfy the minimum length condition will become vortons.
Such loops will ultimately be able to survive as 
vortons if the induced current, and consequently $N$ and $Z$, are 
sufficiently large, such that
\be
\vert NZ \vert^{1/2} \gg {1}.
\ee
Any loop that fails to satisfy this
condition is doomed to lose all its energy and disappear.

The total number density of small loops with length and radial
extension of the order of $L_{\rm min}$, the minimum length for vortons,
will be not much less than the number density
of all closed loops and hence
\be
n\approx \nu\  L_{\rm min}^{-3},
\ee
where $\nu$ is a time-dependent parameter. The typical length scale of
string loops at the transition temperature, 
$L_{\rm min}(\Tx)$, is considerably greater than the relevant thermal 
correlation
length, ${\Tx}^{-1}$, that characterises the local current
fluctuations. 
It is because of this that string loop evolution is
modified after current carrier condensation. Indeed, since 
$L_{\rm min}(\Tx)\gg {\Tx}^{-1}$ and loops present at the time
of the condensation satisfy $L\geq L_{\rm min}(\Tx)$, then
reasonably large values
of the quantum numbers $\vert Z\vert$ and  $N$ build up. 
If $\lambda$ is the wavelength of the flutuation of the carrier field 
then
\be
\vert Z\vert \approx N \approx \left({L\over\lambda}\right)^{1/i} ,
\ee
where $i=1$ in the strictly chiral case and $i=2$ in the more general case,
and $\lambda\approx\Tx^{-1}$. In the above the difference between the
chiral and non-chiral cases arises since there is a random walk effect in
the latter case. Thus, one obtains 
\be
\vert Z\vert \approx N \approx \left({L_{\rm min}(\Tx)\Tx}\right)^{1/i} \gg 1.
\ee
For current condensation during the friction-dominated regime 
this requirement is always satisfied. 

Therefore, the vorton mass density is
\be
\rhv\approx {N \mx \nv} .
\ee
In the friction-dominated regime the string is interacting with the 
surrounding plasma. We can estimate $L_{\rm min}$ in this regime as the
typical length scale below which the microstructure is smoothed \cite
{rob&brandon,CarterDavis:99}. This then gives the quantum number, $N$
\be
N\approx\left({\mP\over\beta\Tx}\right)^{1/{2i}} ,
\label{old 21}
\ee
where $\beta$ is a drag coefficient for the friction-dominated era that is
of order unity. We then obtain the number density of mature vortons
\be
\nv\approx\nua \left({\beta\Tx\over \mP}\right)^{3/2} T^3.
\label{plus 26}
\ee
This gives the resulting mass density of the relic vorton population to be

\be
\rhv\approx \nua \left({\beta\Tx\over\mP}\right)^{3/2 - 1/{2i}}\Tx T^3 \ , 
\label{plus 31}
\ee
where we have assumed that $\mx \sim \Tx$.

\subsection{The Nucleosynthesis Constraint.}

One of the most robust predictions of the standard cosmological model is the
abundances of the light elements that were fabricated during primordial
nucleosynthesis at a temperature $\TN\approx 10^{-4} \GeV$.
In order to preserve this well-established picture, it is necessary that the
energy density in vortons at that time, $\rhv(\TN)$, should have been small
compared with the background energy density in radiation, 
$\rhN\approx\aa\TN^4$, where $\aa$ is the effective number of degrees of 
freedom. Assuming that carrier condensation occurs during the
friction damping regime and that $\aa$ has dropped to a value of order unity
by the time of nucleosynthesis, this gives

\be
\nua\aas^{-1}\left({\beta\Tx\over\mP}\right)^{3/2 - 1/{2i}}\Tx\ll\TN\ ,
\label{old 24}
\ee 
where $\aas$ is the effective number of degrees of freedom at the time
of current condensation and is approximately $10^2$ in the early universe.
If we make the rather conservative assumption that vortons only survive for
a few minutes, which is all
that is needed to reach the nucleosynthesis epoch, we obtain a fairly 
weak restriction on the theory.  

\be 
\Tx\leq 10^8 \GeV
\ee
in the strictly chiral case and
\be
\Tx\leq 10^9 \GeV
\ee
in the more general case, where we have assumed that 
the net efficiency factor $\nua$ and drag factor $\beta$ are of order 
unity. These are the conditions that must be satisfied by the formation 
temperature of {\it cosmic strings that become superconducting immediately}, 
subject to the rather
conservative assumption that the resulting vortons last for at least a few
minutes. We note that both these conditions rule out the formation of such 
strings during any conceivable GUT transition, but are consistent with their 
formation at temperatures close to that of the electroweak symmetry breaking 
transition.

\subsection{The Dark Matter Constraint.}

Let us now consider the rather stronger constraints that can be 
obtained if at
least a substantial fraction of the vortons are sufficiently stable to last
until the present epoch. Indeed one might expect the chiral vortons to be
sufficiently stable \cite{blanco,DDP}. 
It is generally accepted that the virial equilibrium of galaxies and
particularly of clusters of galaxies requires the existence of a cosmological
distribution of ``dark" matter. This matter must have a density 
considerably in excess of the baryonic matter
density, $\rhb\approx 10^{-31}$ gm/cm$^3$. On the other hand, on the same
basis, it is also generally accepted that to be consistent with the
formation of structures such as galaxies it is necessary
that the total amount of this ``dark" matter should not greatly exceed the
critical closure density, namely

\be
\rhc\approx 10^{-29} {\rm gm \ cm^{-3}} \ .
\label{add 15}
\ee
As a function of temperature, the critical density scales like the entropy 
density so that it is given by

\be
\rhc(T)\approx \aa\mc T^3\ ,
\label{plus 35}
\ee
where $\mc$ is a constant mass factor. For comparison with the density of 
vortons that were formed at a scale $\Tx$ we can estimate this to be

\be
\aas \mc\approx 10^{-28}\mP\approx 1\,\hbox{eV}\ .
\label{plus 37}
\ee
The general dark matter constraint is

\be
\Ov \equiv {\rhv\over\rhc}\leq 1\ .
\label{plus 38}
\ee
Inserting our earlier estimate for the vorton density and noting that
$\aas$ is essentially unity in the present epoch, we can derive the
dark matter constraint. This gives

\be
\Tx \leq 10^5\, \GeV\
\ee
in the strictly chiral case and

\be
\Tx \leq 10^7\, \GeV\
\ee
in the non-chiral case, where we have again assumed that the efficiency 
factor and drag coefficent are of order unity. This result is based on the 
assumptions that the vortons in question are stable enough
to survive until the present day. Thus, this constraint is naturally more 
severe than its analogue in the previous section. However, we expect
it to be realistic for the chiral case since such vortons are classically
and quantum mechanically very stable.   
It is to be
remarked that  vortons produced in a phase transition occurring at or near the
limit that has just been derived would give a significant contribution to the
elusive dark matter in the universe. However, if they were produced at the
electroweak scale, then they would constitute such a small dark matter 
fraction, $\Ov\approx 10^{-9}$, that they would be very difficult to detect.

These constraints are very general for long-lived vortons. They raise the 
question of what class of theories give rise to zero modes on cosmic strings.
One such example is that of supersymmetric theories. Indeed a class of 
supersymmetric theories where the gauge symmetry is broken by a 
so-called D-term, gives rise to chiral cosmic strings, and hence the most 
stringent constraints. We consider this in the next section.

\section{A Supersymmetric Model}

We consider supersymmetric versions of the spontaneously broken gauged
$U(1)$ abelian Higgs model \cite{mark1}. These models are related to or are 
simple extensions of those found in reference \cite{Fayet I}. For abelian
theories the gauge symmetry can be broken either by adding a potential
or by a so-called D-term \cite{Fayet II}. The former case also generalises to 
non-abelian theories \cite{mark2}. Here we consider both cases. 

In component form we can write the lagrangian as 

\begin{equation}
\Lag = \Lag_B + \Lag_F + \Lag_Y - U \ ,
\label{nsusyLag}
\end{equation}
with
\begin{eqnarray}
\Lag_B &=& (D^{i\ast}_\mu \phb_i) (D^{i\mu} \phi_i)
		- \frac{1}{4} F^{\mu\nu}F_{\mu\nu} \ , \\
\Lag_F &=& -i\psi_i \sm D^{i\ast}_\mu \psb_i - i\la_i \sm \dmu \lab_i \ , \\
\Lag_Y &=& \frac{ig}{\sqrt{2}} q_i \phb_i \psi_i \la 
 	  - \left(\frac{1}{2}b_{ij} + c_{ijk}\phi_k \right) \psi_i \psi_j 
	  + (\mbox{c.c.}) \ , \\
   U   &=& |F_i|^2 + \frac{1}{2}D^2 ,
\label{Ueqn}
\end{eqnarray}
where $D^i_\mu = \dmu + \frac{1}{2}ig q_i A_\mu$ and 
$F_{\mu\nu} = \dmu A_\nu - \partial_\nu A_\mu$. Here $\phi_i$ are
complex scalar fields of $U(1)$ charge $q_i$, $A_\mu$ is the gauge field
and $g$ is the gauge coupling.
These correspond to the familiar boson fields of the abelian Higgs model. The
fermions, $\psi_i$, $\psb_i$, $\la_i$ and $\lab_i$, are Weyl spinors and the 
complex bosonic fields, $F_i$, and real bosonic field, $D$, are auxiliary 
fields which can be eliminated by their equations of motion.  These are,

\be
F^\ast_i + a_i + b_{ij}\phi_j + c_{ijk}\phi_j\phi_k = 0 \ ,
\ee
\be
D + \kappa + \frac{g}{2} q_i \phb_i \phi_i = 0 \ .
\label{deqn}
\ee
Substituting gives the most general potential to be
\be 
U =  |a_i + b_{ij}\phi_j + c_{ijk}\phi_j\phi_k|^2 
     + \frac{1}{2}\left(\kappa + \frac{g}{2} q_i \phb_i \phi_i \right)^2 \ .
\ee

Now consider spontaneous symmetry breaking in these theories. Considering
the gauge invariance of the underlying superpotential which gave rise to
$U$ implies that $a_i \neq 0$
only if $q_i =0$, $b_{ij} \neq 0$ only if $q_i + q_j =0$ and 
$c_{ijk} \neq 0$ only if $q_i + q_j + q_k=0$. 
The situation is a little more complicated than in non-SUSY theories, since
anomaly cancellation in SUSY theories implies the existence of more than one 
chiral superfield (and hence Higgs field). In order to break the gauge
symmetry, one may either
induce SSB through an appropriate choice 
of superpotential or, in the case of the $U(1)$ gauge
group, one may rely on a non-zero Fayet-Iliopoulos, or D term \cite{Fayet II}.

We shall refer to the theory with superpotential SSB (and, for simplicity, 
zero Fayet-Iliopoulos term) as theory F and
the theory with SSB due to a non-zero Fayet-Iliopoulos term as theory D. 
Since the
implementation of SSB in theory F can be repeated for more general gauge 
groups, we
expect that this theory will be more representative of general defect-forming
theories than theory D for which the mechanism of SSB is specific to the 
$U(1)$ gauge group.

\subsection{Theory F: Vanishing Fayet-Iliopoulos Term}
The simplest model with vanishing Fayet-Iliopoulos term 
($\kappa=0$) and 
spontaneously broken gauge symmetry contains three chiral superfields.
It is not possible to construct such a model with fewer superfields which 
does not
either leave the gauge symmetry unbroken or possess a gauge anomaly.
The fields are two charged fields $\Phi_\pm$, with respective $U(1)$ charges 
$q_\pm = \pm 1$, and a neutral field, $\Phi_0$. 
The potential $U$ is minimised when $F_i=0$ and $D=0$. This occurs when 
$\phi_0=0$, $\phi_+ \phi_- = \eta^2$ and $|\phi_+|^2 = |\phi_-|^2$.
Thus we may write $\phi_\pm = \eta e^{\pm i\al}$, where $\alpha$ is some 
function. We shall now seek the Nielsen-Olesen\cite{NO} solution 
corresponding to an infinite straight cosmic string.  
We proceed in the same manner as for
non-supersymmetric theories. Consider only the bosonic fields (i.e. set the 
fermions to zero) and in cylindrical polar coordinates $(r,\vp, z)$ write 

\begin{eqnarray}
\phi_0 & = & 0 \ , \\
\phi_+ & = & \phi_-^\ast = \eta e^{in\vp}f(r) \ , \\
A_\mu & = & -\frac{2}{g} n \frac{a(r)}{r}\delta_\mu^\vp \ , \\
F_\pm & = & D = 0 \ , \\
F_0 & = & \mu \eta^2 (1 - f(r)^2) \ ,
\label{StringSol}
\end{eqnarray}
so that the $z$-axis is the axis of symmetry of the defect. The profile 
functions, $f(r)$ and $a(r)$, obey 

\begin{equation}
f''+\frac{f'}{r} - n^2\frac{(1-a)^2}{r^2} = \mu^2 \eta^2 (f^2 -1)f \ ,
\label{fEqn}
\end{equation}

\begin{equation}
a''-\frac{a'}{r} = -g^2 \eta^2(1-a)f^2 \ ,
\label{aEqn}
\end{equation}
with boundary conditions 

\bea
f(0)=a(0)=0 \ ,  \\ 
\lim_{r\rightarrow \infty}f(r)=\lim_{r\rightarrow\infty}a(r)=1 \ .
\eea
Note here, in passing, an interesting aspect of topological defects in
SUSY theories. The ground state of the theory is supersymmetric,
but spontaneously breaks the gauge symmetry, while in the core of the defect 
the gauge symmetry is restored but, since $|F_i|^2 \neq 0$ in the core, 
SUSY is spontaneously broken there. 

We have constructed a cosmic string solution in the bosonic sector of the 
theory. Now consider the fermionic sector. In our case this is

\begin{equation}
\Lag_Y = i\frac{g}{\sqrt{2}}
	 \left(\phb_+ \psi_+ - \phb_- \psi_-\right) \la
		 - \mu \left(\phi_0 \psi_+ \psi_- + \phi_+ \psi_0 \psi_- 
		+ \phi_- \psi_0 \psi_+ \right) + {\rm c.c.}
\end{equation}
As with a non-supersymmetric theory, non-trivial zero energy fermion
solutions can exist in the string core. Consider the fermionic ansatz 

\be
\psi_i = \spinU \psi_i(r,\vp) \ ,
\ee
\be
\la = \spinU \la(r,\vp) \ .
\ee
If we can find solutions for the $\psi_i(r,\vp)$ and $\la(r,\vp)$ then, 
following Witten \cite{witten}, we know that solutions of the form

\be
\Psi_i=\psi_i(r,\vp)e^{\chi(z + t)} \ , \quad
 \Lambda=\la(r,\vp)e^{\chi(z + t)} \ ,
\label{witteq}
\ee
with $\chi$ some function, represent left-moving superconducting 
currents flowing along the string at the speed of light. Thus, the
problem of finding the zero modes is reduced to solving for the 
$\psi_i(r,\vp)$ and $\la(r,\vp)$.

The fermion equations of motion derived from \bref{nsusyLag} are four
coupled equations given by

\be
\eth{-i}\left(\dr -\frac{i}{r}\dth \right)\lab - \frac{g}{\sqrt{2}} \eta
		f \left(\eth{in}\psi_- - \eth{-in} \psi_+\right) = 0 \ ,
\label{fermeq1}
\ee
\be
\eth{-i}\left(\dr -\frac{i}{r}\dth \right)\psb_0 + i \mu \eta 
		f \left(\eth{in}\psi_- + \eth{-in} \psi_+\right) = 0 \ ,
\label{fermeq2}
\ee
\be
\eth{-i}\left(\dr -\frac{i}{r}\dth \pm n\frac{a}{r}\right)\psb_\pm +
  \eta f \eth{\mp in}\left(i\mu \psi_0 \pm \frac{g}{\sqrt{2}}\la \right) = 0 
\ .
\label{fermeq3}
\ee
The corresponding equations for the lower fermion components can be obtained 
from those for the upper components by complex conjugation, and 
putting $n \rightarrow -n$. The superconducting current corresponding to this 
solution (like \bref{witteq}, but with $\chi(t-z)$) is right-moving.

We may enumerate the zero modes using an index theorem \cite{index}.
This gives $2n$ independent zero
modes, where $n$ is the winding number of the string. However, in
supersymmetric theories we can calculate them explicitly using SUSY 
transformations. This relates the fermionic components of the superfields to 
the bosonic ones and we may use this to obtain the fermionic solutions in 
terms of the background string fields \cite{mark1}.
\begin{eqnarray}
\la_\al &\rightarrow& \frac{2na'}{gr}i(\sgz)^\beta_\al \xi_\beta \ , \\
(\psi_\pm)_\al &\rightarrow& \sqrt{2} \left(if'\sgr \mp
\frac{n}{r}(1-a)f \sgth\right)_{\al \ald} \xib^\ald \eta \eth{\pm in} \ , \\
(\psi_0)_\al &\rightarrow& \sqrt{2}\mu\eta^2(1-f^2)\xi_\al \ ,
\end{eqnarray}
where we have defined

\begin{eqnarray}
\sgth & = & \mat{0}{-i\eth{-i}}{i\eth{i}}{0} \ , \\
\sgr & = & \mat{0}{\eth{-i}}{\eth{i}}{0} \ .
\end{eqnarray}

Let us choose $\xi_{\alpha}$ so that only one component is nonzero.
Taking $\xi_2 = 0$ and $\xi_1 = -i\delta/(\sqrt{2}\eta)$, where $\delta$
is a complex constant, the fermions become

\begin{eqnarray}
\la_1 & = & \delta\frac{n\sqrt{2}}{g\eta}\frac{a'}{r} \ , \\ 
(\psi_+)_1 & = & \delta^\ast \left[f'+\frac{n}{r}(1-a)f\right]\eth{i(n-1)} 
\ , \\
(\psi_0)_1 & = & -i\delta\mu\eta(1-f^2) \ , \\
(\psi_-)_1 & = & \delta^\ast \left[f'-\frac{n}{r}(1-a)f\right]\eth{-i(n+1)} \ .
\end{eqnarray}
It is these fermion solutions which are responsible for the string
superconductivity.
Similar expressions can be found when $\xi_1 = 0$. It is clear from
these results that the string is not invariant under supersymmetry,
and therefore breaks it. However, since $f'(r), a'(r), 1-a(r)$ and $1-f^2(r)$ 
are all approximately zero outside the string core, the SUSY breaking and
the zero modes are confined to the string. We note that this method gives us 
two zero mode solutions. Thus, for a winding number $1$ we obtain the
full spectrum, whereas for strings of higher winding number only a partial
spectrum is obtained.

The results presented here can be extended to non-abelian gauge theories.
This is done in \cite{mark2}. The results are very similar to those presented
here, so we leave the interested reader to consult the original paper.

\subsection{Theory D: Nonvanishing Fayet-Iliopoulos Term}

Now consider theory D in which there is just one primary charged chiral
superfield involved in the symmetry breaking and a non-zero Fayet-Iliopoulos
term. In order to avoid gauge anomalies, the model must contain other 
charged superfields. These are
coupled to the primary superfield through terms in the superpotential such that
the expectation values of the secondary chiral superfields are dynamically 
zero. The secondary superfields have no effect on SSB and are invariant under
SUSY transformations. Therefore, for the rest 
of this section we shall concentrate on the primary chiral 
superfield which mediates the gauge symmetry breaking. 

Choosing $\kappa = -\frac{1}{2}g\eta^2$, the theory is spontaneously
broken and there exists a string solution obtained from the ansatz
\begin{eqnarray} 
\phi & = & \eta e^{in\vp}f(r) \ , \\
A_\mu & = & -\frac{2}{g} n \frac{a(r)}{r}\delta_\mu^\vp \ , \\
D & = & \frac{1}{2}g\eta^2 (1-f^2) \ , \\
F & = & 0 \ .
\end{eqnarray}
The profile functions $f(r)$ and $a(r)$ then obey the first order equations

\begin{equation}
f' = n\frac{(1-a)}{r}f,
\label{fEqnII}
\end{equation}

\begin{equation}
n\frac{a'}{r} = \frac{1}{4}g^2 \eta^2 (1-f^2).
\label{aEqnII}
\end{equation}

Now consider the fermionic sector of the theory and perform a SUSY
transformation. This gives

\begin{eqnarray}
\la_\al &\rightarrow&  \frac{1}{2}g\eta^2(1-f^2)i(I+\sgz)_\al^\beta\xi_\beta,\\
\psi_\al &\rightarrow& 
\sqrt{2}\frac{n}{r}(1-a)f(i\sgr-\sgth)_{\al \ald} \xib^\ald \eta \eth{in}.  
\end{eqnarray}
If $\xi_1=0$ both these expressions are
zero. The same is true of all higher order terms, and so the string is
invariant under the corresponding transformation. For other $\xi$,
taking $\xi_1= -i\delta/\eta$ gives
\bea 
\la_1  &=& \delta g \eta(1-f^2), \\
\psi_1 &=& 2\sqrt{2} \delta^\ast \frac{n}{r}(1-a)f \eth{i(n-1)}. 
\eea
Thus supersymmetry is only half broken inside the string. Vortices with
supersymmetry half broken in the string core also arise in the theories
considered in \cite{ana}.
This is in
contrast to theory F which fully breaks supersymmetry in the string
core. The theories also differ in that theory D's zero
modes will only travel in one direction, while the zero modes of theory F 
(which has twice as many) travel in both directions. Thus the D theory
has chiral zero modes and is subject to the constraints derived in
the previous section for chiral vortons.
In both theories the
zero modes and SUSY breaking are confined to the string core.

Thus, a necessary feature of cosmic strings in SUSY theories is that 
supersymmetry is broken in the string core and the resulting strings have
fermion zero modes. As a consequence, cosmic strings arising in SUSY theories
are automatically current-carrying. As discussed in the previous section,
the presence of vortons puts severe constraints on the underlying 
theory since the density of vortons could overclose the universe if they
are stable enough to survive to the present time. 
If they only live for a few minutes then the vorton density could affect
nucleosynthesis. 

\subsection{Soft Susy Breaking}
Supersymmetry is not observed in nature. Hence, it must be broken. 
Supersymmetry breaking is achieved by adding soft SUSY breaking terms which do
not induce quadratic divergences.  

In a general model, one may obtain soft SUSY breaking terms by the 
following prescription.

\begin{enumerate}
\item Add arbitrary mass terms for all scalar particles to the scalar 
potential.

\item Add all trilinear scalar terms in the superpotential, plus
their hermitian conjugates, to the scalar potential with arbitrary coupling.

\item Add mass terms for the gauginos to the Lagrangian density.
\end{enumerate}

Since the techniques we have used are strictly valid 
only when SUSY is exact, it is necessary to investigate the effect of these 
soft terms on the fermionic zero modes we have identified.

As we have already commented, the existence of the zero modes can be seen as 
a consequence of an index theorem~\cite{index}. The index is
insensitive to the size and exact form of the Yukawa couplings, as
long as they are regular for small $r$ and tend to a constant at
large $r$. In fact, the existence of zero modes relies only on
the existence of the appropriate Yukawa couplings and that they have the 
correct $\vp$-dependence. Thus there can only be a change in the number of zero
modes if the soft breaking terms induce specific new Yukawa couplings
in the theory and it is this that we must check for. Further, it was
conjectured in~\cite{index} that the destruction of a zero mode occurs only 
when the relevant fermion mixes with another massless fermion. 

We have examined each of our theories with respect to this criterion and list 
the results below.

\subsubsection{Theory-F}

The trilinear and mass terms that arise from soft SUSY breaking
are
\be
m_0^2 |\phi_0|^2 + m_-^2 |\phi_-|^2 + m_+^2 |\phi_+|^2 
 + \mu M \phi_0\phi_+\phi_-.
\ee
The derivative of the scalar potential with respect to $\phi_0^\ast$ becomes
\be
\phi_0 (\mu^2|\phi_+|^2 + \mu^2|\phi_-|^2 + m_0^2) + \mu M(\phi_+ \phi_-)^\ast.
\ee
This will be zero at a minimum, and so $\phi_0 \ne 0$ only if $M \ne 0$.

New Higgs mass terms will alter the values of $\phi_+$ and $\phi_-$ slightly, 
but will not produce any new Yukawa terms. Thus these soft
SUSY-breaking terms have no effect on the existence of the zero modes.

However, the presence of the trilinear term gives $\phi_0$ a non-zero 
expectation value, which gives a Yukawa term coupling the $\psi_+$ and $\psi_-$
fields. This destroys all the zero modes in the theory since the left-
and right-moving zero modes mix. 
For completeness note that a gaugino mass term also mixes the left and right 
zero modes, aiding in their destruction. Thus in this case there is a 
possibility of the fermion zero modes being destroyed by supersymmetry 
breaking and evading the vorton bounds of the previous section.
  
\subsubsection{Theory-D}

The $U(1)$ theory with gauge symmetry broken via a Fayet-Iliopoulos term and 
no superpotential is simpler to analyse. New Higgs mass terms 
have no effect, as in the above  case, and there are no trilinear
terms. Further, although the gaugino mass  terms affect the form of
the zero mode solutions, they do not affect their  existence, and so,
in theory-$D$, the zero modes remain even after SUSY breaking. For 
this class of theories, the strings remain current-carrying,
have a vorton problem, and are therefore in conflict with cosmology unless
subsequent phase transistions destroy the zero modes.

\section{Microwave Background Constraints}

In previous lectures \cite{mairi} we have seen the cosmic string 
predictions for the cosmic microwave background. However, the 
microwave background can also be used to constrain theories in another
respect. 
Cosmological models with non-thermal photon production occurring after
a redshift of about $10^6$ are strongly constrained by the precision 
measurements \cite{FIRAS}
of the black-body spectrum of the Cosmic Microwave Background (CMB).
Energy release (either into photons or electrons and other electromagnetically
interacting particles with rather general spectra) for redshift
in the range between $10^5$ and $3\times10^6$ would lead to a Bose--Einstein
distortion of the thermal spectra characterised by a chemical potential
\cite{FIRAS1}. Energy release at later times produces a distortion in
the Comptonised spectrum. The bottom line is that 
the fraction of photons produced after the above redshift must be
smaller than $7 \times 10^{-5}$.
This can be used to constrain any theory with decaying topological
defects. For example, theories which give rise to embedded defects 
\cite{tanmay&ana}
or theories resulting in vortons which are not stable to the present
time can be constrained in this way \cite{rob&brandon2}.

Some types of embedded defects can be stabilised by plasma 
effects in
the early universe when some of the fields are electromagnetically
charged via interaction with the surrounding photons \cite{Nagasawa}. Since
the photons fall out of thermal equilibrium at the time $t_{\rm ls}$ of
last scattering, the stabilization forces will disappear at this time
and the embedded defects will decay, emitting a certain fraction of
their energy as non-thermal photons. They can thus be constrained
by the FIRAS results.

To illustrate the mechanism that renders some embedded defects stable
in the early Universe, we shall consider a model~\cite{Carter:2002te} 
in which the order
parameter consists of four real scalar fields with a standard
symmetry breaking potential symmetric in the four fields. We will
assume that two of the fields are electrically charged and the other
two neutral, as is the case in the standard electroweak theory and
in the Sigma model description of low energy QCD in the limit of
vanishing pion mass.

In an electromagnetic plasma, the interactions with the photon bath
will lead to an effective potential for the order parameter which
breaks the symmetry between the charged and the neutral fields, while
preserving the symmetry within the neutral scalar field sector. The
potential is lifted more in the direction of the charged scalar fields.
Thus, the vacuum manifold becomes $S^1$, giving rise to cosmic
string solutions of the full field equations which look like the
standard cosmic string configuration of the neutral fields
with the charged scalar fields set to zero.

The thermal averaging implicit in the above analysis breaks down
after the time of last scattering, and thus it is expected that
the confining potential will disappear and the embedded strings will
decay, emitting a fraction $\beta$ (which is expected to be smaller but
of the order of $1$) of its energy as photons. Since the topological
defects are out-of-equilibrium objects, the photons produced by
their decay will lead to spectral distortions of the thermal CMB.

Let us now look at this stabilization mechanism in more detail.
The Higgs sector of our model is described by the following
Lagrangian
\be \label{01} 
{\cal L} \, = \, {1\over 2}
\big(D_{\mu}\chi^\ast D^{\mu}\chi +\dmu\phi^\ast{\partial^{\mu}\phi}\big) 
- V\, ,
\ee
with the symmetry breaking potential
\be \label{02} 
V \, = \, {\lambda\over 4}\big(\chi\,^\ast\!\chi+ 
\phi\,^\ast\!\phi-\eta^2\big)^2  \, ,
\ee
and with gauge-covariant derivative
\be \label{03}
D_{\mu}\chi \, = \, \dmu\chi+ie A_\mu\chi \, ,
\ee
where $e$ is the fundamental charge and $A_{\mu}$ is the gauge field
of electromagnetism. The field $\phi$ denotes the neutral Higgs
doublet, which gives rise to the string solution and 
the field $\chi$ is the charged Higgs doublet. Using the
language of the low-energy Sigma model of QCD we can write
\be \label{04} 
\chi \, = \, \pi_1+ i\pi_2\, ,\hskip 0.7 cm \phi \, = \, \pi_3+i\pi_0
\, .
\ee 

The finite temperature corrections to the potential of the theory given
by (\ref{01}) were worked out in detail in \cite{Carter:2002te}. 
Keeping only the contributions to the finite temperature effective
potential $V_T$ quadratic in the temperature $T$ coming from the
photon bath we obtain
\begin{eqnarray} \label{05}  
V_T - V \, &=& \, {{T^2} \over 2} e^2\Big({1\over 6}
A_\mu A^{\mu}+ \pi\, \chi\,^\ast\!\chi\Big)\nonumber\\ 
 &+& {T^2}{\lambda\over 4}\big(\chi\,^\ast\!\chi+
\phi\,^\ast\!\phi - {2\over 3}\eta^2) \, .
\end{eqnarray}
The second term in the first line is responsible for the breaking of
the degeneracy in the potential between the charged and neutral scalar
fields. Equation (\ref{05}) describes a potential which is lifted by
different amounts in the neutral and charged scalar field directions
compared to the zero temperature potential. For temperatures below
the critical temperature $T_c = \sqrt{2} \eta$, the space of lowest
energy states forms a manifold $S^1$ consisting of configurations
with $\chi = 0$ and $\phi^{\ast} \phi = \eta_{_T}^2$, where 
$\eta_{_T}$ is given by
\be \label{06} 
\eta_{_T}^{2} \, = \, \eta^2- {{T^2} \over 2}\, .
\ee 
Thus, there is a static string solution consisting of a $U(1)$ cosmic
string in the $\phi$ variables with $\chi = 0$, the {\it embedded string}.

The embedded string is stable in the temperature range immediately below
the critical temperature for which the curvature of the potential in
the $\chi$ direction is positive at $\chi = 0$. This is the case for
\be \label{08} 
\sqrt{2}\Big(1+{2\pi e^2\over\lambda}\Big)^{-1/2} \, < \, {T\over\eta} \,
< \, \sqrt{2}\, .
\ee

When the temperature drops below the lower limit given in (\ref{08}), 
the string does not disappear, but rather undergoes a core phase 
transition \cite{Axen1,Axen2} 
in which the charged field $\chi$ 
acquires a non-vanishing value $\chi_T$ in
the string core which lowers the potential energy density in the core.
What results is an asymmetric vortex defect which, as 
realized in \cite{Carter:2002te}, will generically be superconducting,
the superconductivity being induced by the phase gradient of $\chi$ in
the string core. This gave rise to the so-called drum vortons, whose 
properties are outside the scope of these lectures.
The value of $|\chi_{_T}|$ is given by
\be \label{09} 
\chi\,^\ast\!\chi \, = \, \eta^2-{T^2 \over 2}\Big(
1+{2\pi e^2\over\lambda}\Big)\, .
\ee

Since the asymmetric vortices are superconducting, string loops will
generically form vortons \cite{vorton}. The strings acquire their
current at the time $T_{_{\rm Q}}$ of the core phase transition, 
i.e. (see (\ref{08}))
\be \label{14}
T_{_{\rm Q}} \, = \, \sqrt{2} \Big(1+{2\pi e^2\over\lambda}\Big)^{-1/2} 
\eta \, .
\ee
Since this temperature is only slightly lower than the temperature
at which the strings initially form, the string network will still
be in the friction-dominated phase when the current condensation
occurs. 

In the following, we will use the knowledge of the energy contained
in embedded vortices, in particular those of asymmetric core nature,
to determine the cosmological constraints.

\subsection{Constraints}

Cosmic vortices (and other defects) that decay into 
(in part) photons at some 
temperature $T_{\rm d}$ corresponding to a redshift of 
less than $10^6$
will produce spectral distortions of the CMB and will thus be strongly
constrained by the COBE/FIRAS data~\cite{FIRAS}. The constraint
on the non-thermal fractional energy density production in photons is
\be \label{fir}
{{\delta \rho_{\gamma}} \over {\rho_{\gamma}}} \, \lta \, 7 \times 10^{-5}.
\ee 
This constraint arises from the limits to the Compton $y$ parameter and 
chemical potential, which measures the spectral distortion. Any
defects which decay after a redshift of $10^6$ will be subject to these FIRAS 
constraints. Thus, the following considerations will apply to
stabilized embedded defects and to decaying topological defects (such a
decay might be induced by a late time phase transition).

The specific bounds on defect models will depend on the type of defect,
on the density of defects and on the specifics of the decay. We will
focus on cosmic strings, both of topological and of embedded type.
First, we consider strings which are in their scaling regime at the time
of decay. This applies, for example, to embedded vortices formed at early
times and in which superconductivity is absent or too weak to produce
vortons. If the strings are formed later than some
critical time, the string network will
not yet be scaling and the dynamics will still be friction-dominated.
Since the density of strings is higher in this phase relative to a
string scaling configuration, the bounds in this case are different.
This is the second case we treat. Finally, we analyze the case of a gas 
of vortons produced by the superconducting string loops. 

\subsubsection{Scaling Cosmic Strings}

First let us examine the case of cosmic strings in the scaling regime. 
The density of strings at time $t$ is
\be \label{17}
\rho_{\rm s} \, = \, \nu {{\eta^2} \over {t^2}} \, ,
\ee
where the constant $\nu$ determines the number of long string segments per
Hubble volume, and whose value is $\nu \sim 10$ (see \cite{VS} and
references therein).

Let us first consider strings decaying at the time $t_{\rm ls}$ of last
scattering. From (\ref{17}) it follows that the density of strings at 
last scattering is
\be \label{18}
\rho_{\rm s} \, = \, \nu {{32\pi^3}\over {45}} {{z_{\rm eq}}
\over{z_{\rm ls}}} {{\eta^2}\over{\mP^2}} T_{\rm ls}^{\, 4} \, ,
\ee
where $z_{\rm eq}$ is the redshift at the time of equal matter and radiation.
If a fraction $\beta$ of the energy of the strings goes directly or indirectly
into photons, then the COBE/FIRAS constraint (\ref{fir}) becomes
\be \label{19}
{{\beta\rho_{\rm s}}\over{\rho}} \, = \, 
\beta\nu {{32\pi} \over 3} \left(\eta \over \mP\right)^2 
{{z_{\rm eq}} \over {z_{\rm ls}}} \, \lta \, 7 \times 10^{-5}.
\ee
Assuming that $\beta$ is of order $1$ then this results in the constraint 
\be \label{21}
\eta \, \lta \, \nu^{-1/2} 10^{16} {\rm GeV}.
\ee
Thus, for values of $\nu$ in the range indicated by present cosmic string
simulations, the COBE/FIRAS constraint severely constrains decaying
cosmic string models with a symmetry breaking scale given by the scale of
Grand Unification, though all cosmic string models with
this scale of symmetry breaking are already constrained \cite{mairi}.

Consider now strings decaying at a redshift $z_{\rm d}$ larger than 
$z_{\rm eq}$ (but smaller than $10^6$). In this case, the COBE/FIRAS 
constraint yields a result analogous to above, but modified by a factor 
of $(z_{\rm eq} / z_{\rm ls})^{1/2}$, resulting in a weaker bound 
by a
factor of about 3 than the bound given in (\ref{21}). Conversely, if
the strings decay after $t_{\rm ls}$, the bound is stronger by a factor
of $(z_{\rm d} / z_{\rm ls})^{1/2}$.

\subsubsection{Friction Dominated Strings}

If however the string network is friction dominated at the time of 
decay then the density of strings is much greater than that in the
scaling regime. However, 
the friction domination condition will be satisfied at the time 
when the strings decay (assumed to be $t_{\rm ls}$) whenever $T_{\star}$ 
is less than $T_{\rm ls}$, i.e. less than $10^{-13}$ GeV. This gives
\be \label{23}
{\eta\over\mP} \lta 10^{-16},
\ee
i.e. $\eta \lta 10^3$ GeV. In this case, though, strings that are
in the friction epoch are not constrained by the COBE/FIRAS
data, unless they lead to vortons.

\subsubsection{Vortons}

In section 3 we saw that cosmic string theories giving rise to vortons
are subject to stringent constraints if they are absolutely stable and
to slightly less stringent constraints if the vortons survive until
the time of nucleosynthesis. Here we consider the additional 
constraint arising from the limit (\ref{fir}) provided by the COBE/FIRAS 
data~\cite{FIRAS}  in cases for which vortons decay between a redshift 
of $10^6$ and today, thus giving rise to decay products that would 
produce observable distortions of the black body spectrum.

If the string is formed at temperature $T_{_{\rm X}}$ then the
vorton density at temperature $T$ is given by \cite{rob&brandon}
\be \label{26}
\rhv \, = \, {\tilde \nu} \Bigl({T_{_{\rm X}}\over\mP}\Bigr)^{5/4}
T_{_{\rm X}} T^3 \, ,
\ee 
where ${\tilde \nu}$ is a constant of the (rough) order of $1$. This
result holds both in the radiation and matter dominated phases.

As before, we denote the temperature at which the vortons decay by 
$T_{\rm d}$. If the vortons emit a fraction $\beta$ of their energy as 
photons, then the fractional photon energy density input from vorton 
decay is
\be \label{27}
{{\Delta \rho_\gamma(T_{\rm d})} \over {\rho_\gamma(T_{\rm d})}} 
\, = \,   \kappa \Bigl({T_{_{\rm X}}\over\mP}\Bigr)^{5/4}
{T_{_{\rm X}} \over T_{\rm d}} \, ,
\ee
in which  the constant $\kappa$ is given by
\be \label{28}
\kappa \, = \, {{30} \over \pi^2 g_*(T_{\rm d})} \beta {\tilde \nu} \, .
\ee
The COBE/FIRAS constraint (\ref{fir}) thus becomes
\be \label{29}
\kappa \Bigl({T_{_{\rm X}}\over\mP}\Bigr)^{5/4}
{T_{_{\rm Q}} \over T_{\rm d}} \, < \, 7 \times 10^{-5} \, .
\ee
Assuming that the decay occurs at the time of last scattering and  
using the estimate $\kappa \sim 1$, the general constraint (\ref{29})
leads to
\be \label{30}
T_{_{\rm X}}\, \lta \, 10^5 {\rm GeV} \, .
\ee

We have considered cosmological constraints on models with decaying
topological defects which result from demanding that the photons
produced in the decay do not lead to spectral distortions of the CMB
in excess of the observational limits from the COBE/FIRAS experiment.
The strongest limits arise for theories leading to decaying vortons.
Any theory giving rise to vortons resulting from a string forming phase 
transition above
$10^5$ GeV, and which decayed from a redshift of $10^6$ to today, would
produce a too large spectral distortion as measured by the 
Compton $y$ parameter, and thus be 
ruled out. The above constraint applies both to vortons resulting from
topological strings or those resulting from embedded strings which have
become stabilised by plasma processes. These constraints are
in fact stronger than constraints on vorton models requiring 
compatibility with nucleosynthesis and with the dark matter abundance
limits \cite{rob&brandon} and are comparable to those obtained in
the chiral case \cite{CarterDavis:99}.

Since many types of embedded strings are stabilized by interactions
with the electromagnetic plasma, undergo core phase transitions and
become superconducting, thus yielding vortons \cite{Carter:2002te}
which decay at the time of last scattering,
our constraints are very important for theories with embedded defects.

Our analysis also gives rise to constraints on theories with 
topological cosmic strings with GUT scale
symmetry breaking scale, provided they decay in the relevant
redshift interval.

\section{Dilatonic Cosmic Strings}

Superstring theory predicts the existence of light, gauge-neutral 
fields (the dilaton and moduli) with gravitational strength 
couplings to ordinary matter. In addition it has an axionic symmetry.
In the effective lagrangian a combination of the dilaton and axion
symmetry is broken to give rise to cosmic string solutions. These cosmic
strings have novel solutions in that there are two energy scales and
oscillating string loops copiously emit dilatons in addition to gravitons.
Since the cosmic strings arise from a supersymmetric theory, there will
also be zero modes in the string core.

Due to the weak couplings with ordinary matter the dilaton is rather
long-lived, posing problems for cosmology. It was realised \cite{DV}
that dilatons are copiously produced by oscillating cosmic string loops,
resulting in constraints on the dilaton mass and scale of symmetry 
breaking. In this section we briefly summarise their results. Since
\cite{DV} only considered the dilaton field and did not take the 
axionic symmetry into account we will also extend their results to 
this case.

\subsection{Dilaton Emission from Cosmic String Loops}

In \cite{DV}, see also \cite{PS}, the dilaton emission from cosmic
string loops was calculated in an analogous way to graviton emission
(see \cite{VS}). That is the energy loss from an oscillating, periodic
source was estimated. The scaling distribution for the string loops
was taken. For the case considered by \cite{DV} the dominant contribution
arose from loops smaller than the critical length,
\be
L_c = {4\pi\over{m_s}},
\ee
where $m_s$ is the mass of the dilaton. Here they found that the total
number of dilatons produced by a decaying loop is
\be 
N = \Gamma\alpha^2G^2\mu^3t_f^2,
\ee
where $\Gamma$ is a numerical coefficient which depends on the loop
trajectory, but is expected to be of order $10^2$, $\alpha$ is a 
dimension-less quantity which measures the strength of the coupling
of the dilaton field to the cosmic strings, and is expected to be
of order $1$, and $t_f$ is the time the loop decays.
The total number of dilatons produced by the string network is
\be
Y(t_f) = {n_s N\over{s(t_f)}}, 
\ee
where $n_s$ is the number density of string loops, calculated using
the scaling distribution and $s(t_f)$ is the entropy density of
the universe at the time of loop decay. This gives,
\be 
Y(t_f) = {{\Gamma\alpha^2}(G\mu)^2({\mP}\,t_f)^{1/2}\over{\aa^{1/4}}}.
\ee

The dilaton constraints are sensitive to its lifetime, which are determined
by its mass and couplings, and is
\be
\tau_s \approx {{4\mP^2}\over{N_{_F}m_s^3}},
\ee
where $N_{_F}$ is the number of gauge bosons with masses less than
$m_s$ and factors of order unity have been ignored. The scale of symmetry
breaking and dilaton mass can then be bound by assuming that dilaton
production does not affect standard nucleosynthesis. It was found that,
for 
\be 
m_s = 10^3 \GeV,
\ee
\be
\eta \le 10^{11} \GeV,
\ee
or for
\be
\eta = 10^{16} \GeV,
\ee
then 
\be
m_s \ge 10^5 \GeV.
\ee

Thus the favoured values of a GUT scale symmetry breaking, $\eta=10^{16} \GeV$,
and dilaton mass given by the scale of supersymmetry breaking,
$m_s = 10^3 \GeV$, are incompatible with observation.
 
\subsection{A Dilaton Model with Axionic Symmetry}

In the above calculation a simple model was used where the complexities
of the superstring effective lagrangian were ignored. I have addressed
this in work in progress with Pierre Binetruy and Stephen Davis.
Here I present our {\it preliminary} results. 
The superstring effective action results in a model with a pseudo-anomolous
$U(1)$ local symmetry and dilaton field. The real part of the dilaton
field defines the gauge coupling, $1/g^2 = \sR$. The bosonic part is discussed
in \cite{ax1}. In the supersymmetric case the gauge symmetry is broken
by a combination of $F$ and $D$ terms, resulting in two distinct mass
scales. The heavier scale, $m_D$ is 
related to the compactification scale in string theory, so should be
close to the Planck scale, whilst $m_F$ is the dilaton scale and should
be related to the supersymmetry breaking scale, thought to be around
$10^3$ GeV.

We look for cosmic string solutions of the form
\bea
\phi &=& \eta f(r) e^{in\vp}, \label{ansst} \\
A_\theta &=&  n \frac{v(r)}{r},  \\
s &=& \frac{\dGS}{\eta^2 \g(r)^2} + 2 i n \dGS \vp, \\
\phi_Z &=& 0, \label{ansend}
\eea
where $\dGS$ is the Green-Schwarz parameter in string theory (see \cite{ax1}
for further details) and $\phi_Z$ is a scalar fieldof charge $q_Z$.
As $r$ approaches infinity the functions $f$, $\g$ and $v$ all tend to 1, 
and we allow spatial variations of the dilaton. 

These profile functions are then inserted into the equations of motion,
which were solved numerically. As expected, two distinct length scales were 
found, resulting in the string having an inner core of radius 
$r_D \sim m_D^{-1}$ ($\rho_D \sim 1$) in which $v < 1$ and $f \neq \g$, 
so $D \neq 0$. Around that region there is an outer core in which 
$v \approx 1$ and $f \approx \g$ but $f,\g < 1$. This
region is of radius $r_F \sim m_F^{-1}$, which can be far greater than
$r_D$ even for moderate values of the parameters.

An approximate analytic solution to the field equations was also found. 
Inside the inner core ($\rho <1$) of
the string the kinetic terms dominate the field equations. Imposing the
boundary conditions $\g=f=v=0$ at $\rho =0$ and $f \approx \g$, 
$v \approx 1$ at $\rho =1$ gives the approximate solution for $\rho <1$
\be
f \approx A \rho^{|n|},
\label{fsmall}
\ee
\be
\g \approx \frac{A}{\sqrt{1-2|n|\eta^2 A^2\log\rho }},
\label{gsmall}
\ee
\be
v \approx \frac{\rho^2}{1-2|n|\eta^2 A^2\log\rho }.
\label{vsmall}
\ee
Note that there is some analogy to the situation in the $A$ phase of $He^3$ 
where there is a hard inner core and a soft outer core \cite{grisha};
a similar situation arises in some supersymmetric theories with flat
directions \cite{lowH}.

If $m_D \gg m_F$ then an approximate solution for the outer core is also
needed. Here the auxiliary field $D \approx 0$ so $f \approx \g$, and we 
obtain
\be
f \approx \g \approx A+ \frac{1-A}{\log(m_D/m_F)} \log \rho,
\ee
for $1 < \rho < m_D/m_F$ and $v=1$. For $\rho > \rho_F$, $f=\g=v=1$.
The constant $A$ is determined by taking $\g'$ to be continuous at
$\rho=1$. 
Note that if $m_D \sim m_F$, there is no outer core. Then $|\phi|$ and $\sR$ 
take their vacuum expectation values for $\rho >1$, so $v=\g=f=1$ and thus 
$A=1$. The energy per unit length is found to be $\mu \sim |n| \eta^2$, with 
logarithmic dependence on $m_D/m_F$. 

Since the theory is supersymmetric there is also a fermionic lagrangian.
From this one can investigate the fermion zero mode solutions. We find
that there is a rich spectrum in this model, found by analogous methods
to those used in the previous sections. In particular these were found
by supersymmetry transformations on the boson fields and also by
use of the index theorem \cite{index}. We find
(for $q_Z<0$) there are $|n|(1 - q_Z)$ zero modes, all of which are
left (right) movers if $n>0$ ($n<0$). Thus these strings are again
current-carrying strings and since the fermions are either left- or
right-movers, the current is chiral. However, the existence of the
two distinct mass scales, $m_D$ and $m_F$, means that we cannot 
automatically assume that the vorton constraints derived previously
will apply. 

\subsubsection{Constraints}

In our case a stable string network is only formed at $T \sim m_F$,
which can be far less than the energy scale of the strings, $\eta$, which is
determined by $m_D$. This is an unusual feature of our model. If
\be
m_F \le \frac{m_D^2}{\mP},
\ee
the friction domination regime will be absent. In this case the
string network may well be formed in the scaling regime.

The unusual form of the axion strings in this model does provide an
alternative solution to the vorton problem. Typically the radius of a
stable loop will be of order $10^2 m_D^{-1}$. If $m_F \le 10^{-3} m_D$
the outer cores of each side of a potential vorton will overlap. It is then
energetically favourable for the proto-vorton loop to decay, releasing
the parent particles as radiation. This will avoid the vorton bounds,
but different bounds arising from constraints on dilaton production will
apply instead. 

We can recalculate the vorton bounds taking $m_F = 10^{-2}m_D$. We find
that there were no constraints on the model arising from the nucleosynthesis
bound. If the vortons were stable enough to live to today, we found

\be
m_F \le \left[ \mP m_c \left(
\frac{\Gamma 10^2 m_F}{\sqrt{g_*} \mP}\right)^{-1/3} \right]^{1/2},
\ee
($m_c \sim 10^{-9} \GeV$ is the closure mass). 
Thus for both $\Gamma, g_* \sim 10^2$ and $m_D = 10^2m_F$ we obtain 
the following bound
\be
m_F \le 10^6 \GeV.
\ee

This is less tight than the constraint we found in section $3$. If we took
$m_F \sim m_D$ then there would only have one scale and we recover the
results of section $3$. Since $m_F$ is essentially the mass of the
dilaton, the usual choice of $10^3 \GeV$ would evade the vorton
constraints. 

We should now check our string loops for dilaton emission. Here the
calculation is similar, but more complicated than that presented above.
This is because the two scales result in the string loops being much
larger than those considered in \cite{DV}. Using the favoured value
of $10^3 \GeV$ for the dilaton mass we find  
\be
\eta \le 10^{14} \GeV.
\ee
Whilst this is still much less than the GUT scale it is higher than 
previous results, opening up the dilaton window in superstring models. 
We also consider direct dilaton emission from loop decay, but this gave
a similar bound.

The final constraint on the theory we need to consider comes from structure
formation. 
The string network will give rise to density perturbations in the microwave
background. Experimentally these are detected at the level of about
$10^{-6}$ \cite{mairi}. This constrains $g^2 \dGS \le 10^{-6}$,
which is rather less than the usual string theory values.

\section{Discussion}

In these lectures we have considered the constraints on particle
physics models arising from cosmic defects. We have seen that
the presence of a current radically changes the cosmology of
a cosmic string network, resulting in the formation of stable
loops or vortons. By calculating the vorton density we were able
to constrain the underlying theory in two ways. If vortons survive
for only a few minutes, we demanded that the universe be radiation
dominated at nucleosynthesis. If the vortons survive until the present
time then we require that they do not overclose the universe. This 
puts stringent constraints on the theory, in particular in the chiral
case. 

We showed that cosmic strings arising in supersymmetric theories
were current-carrying via fermion zero modes. In some cases the
zero modes survived supersymmetry breaking, so the vorton
constraints applied. Chiral zero modes were obtained when the
gauge symmetry was broken with a $D$ term; in this case the 
zero modes survived supersymmetry breaking.

Cosmological constraints on models with decaying defects were obtained
by demanding that the photons
produced in the decay do not lead to spectral distortions of the CMB
in excess of the observational limits from the COBE/FIRAS experiment.
The strongest limits arise for theories leading to decaying vortons.
Any theory giving rise to vortons resulting from a string forming phase 
transition above
$10^5$ GeV, and which decayed from a redshift of $10^6$ to today, would
produce a too large spectral distortion and thus be 
ruled out. This constraint applies both to vortons resulting from
topological strings or those resulting from embedded strings which have
become stabilised by plasma processes. This constraint is comparable
to that found for chiral vortons and stronger than the other cases.

Finally we showed that cosmic strings arising in superstring theory
were subject to constraints from dilaton production. We showed that
in such models there are two distinct scales resulting in cosmic strings
with an inner and outer core. This raises the possibility that such 
theories could evade vortons constraints, though they would still be subject
to constraints arising from the overproduction of dilatons.

\noindent{\bf Acknowledgments}

These lectures were presented at the Summer School on {\it
Patterns of Symmetry Breaking}, supported by NATO and the European
Science Foundation Programme {\it Cosmology in the Laboratory}.
I would like to thank the organisers for the opportunity
to deliver these lectures, and creating a friendly and stimulating
enviroment. I would also like to thank my collaborators for the fun
we have had over the years developing the ideas presented here, 
and my colleagues who read preliminary
versions of these lectures. This work is supported in part by PPARC
and the ESF.

\end{document}